\documentclass[aps,prl,
              reprint,
              amsmath,amssymb,amsfonts,groupedaddress,
              longbibliography,
              superscriptaddress,floatfix]{revtex4-1}

\usepackage{graphicx}
\usepackage{dcolumn}
\usepackage{bm}
\usepackage{units}

\usepackage{color}

\usepackage[breaklinks=true]{hyperref}
\usepackage{hyperref}
  \hypersetup{
  colorlinks=true,
  linkcolor=red,
  citecolor=green,
  filecolor=black,
  menucolor=black,
  urlcolor=blue
  }
 
\usepackage{breakurl}

\newcommand{\sdist}{\kern 0.20em}
\renewcommand{\eqref}[1]{Eq.\sdist(\ref{#1})}

\allowdisplaybreaks

\usepackage{upgreek}

\begin{document}

\title{On-shot diagnostic of electron beam-laser pulse interaction based on\\stochastic quantum radiation reaction}

\author{Matteo Tamburini}\email{matteo.tamburini@mpi-hd.mpg.de}
\affiliation{Max-Planck-Institut f\"ur Kernphysik, Saupfercheckweg 1, D-69117 Heidelberg, Germany}

\date{\today}

\begin{abstract}
Ultrarelativistic electron beam-laser pulse scattering experiments are the workhorse for the investigation of QED and of possible signatures of new physics in the still largely unexplored strong-field regime. However, shot-to-shot fluctuations both of the electron beam and of the laser pulse parameters render it difficult to discern the dynamics of the interaction. Consequently, the possibility of benchmarking theoretical predictions against experimental results, which is essential for validating theoretical models, is severely limited. Here we show that the stochastic nature of quantum emission events provides a unique route to the on-shot diagnostic of the electron beam-laser pulse interaction, therefore paving the way for accurate measurements of strong-field QED effects.
\end{abstract}

\maketitle

Stimulated by the advances in the development of high-intensity lasers~\cite{yanovskyOE08, korzhimanovPU11, cheriauxACP12, HPL:9530868} and the prospect of 10-PW class laser facilities~\cite{eliURL, xcelsURL} there is steadily growing interest both in the theoretical and in the experimental
investigation of QED and of new possible exotic processes in the presence of strong background electromagnetic fields~\cite{marklundRMP06, mourouRMP06, ehlotzkyRPP09, bulanovPPCF11, dipiazzaRMP12}. Strong-field QED (SFQED) processes are primarily characterized by the quantum parameter $\chi = F^*/F_{\text{cr}}$, where $F^*$ denotes the electric field in the rest frame of a particle and $F_{\text{cr}}=m_e^2 c^3/|e| \hbar \approx 1.3\times 10^{18}\,\text{V/m}$ is the QED critical (Schwinger) field. Thus, due to the Lorentz boost of the electric field $F^*$ experienced by the beam particles, the study of SFQED effects is greatly facilitated by considering ultrarelativistic electron beams colliding with intense laser pulses. A prominent example of this class of experiments is the investigation of the influence of the energy losses associated to the emission of radiation on the dynamics of the emitting particle, i.e., the so-called radiation reaction (RR) effects~\cite{kogaPoP05, dipiazzaLMP08, dipiazzaPRL09, bulanovPRE11, thomasPRX12, neitzPRL13, kravetsPRE13, tamburiniPRE14, greenPRL14, liPRL14, vranicPRL14, blackburnPRL14, wangPOP15, heinzlPRE15, vranicNJP16, harveyPRL17, nielPRE18, colePRX18, poderPRX18, liPRA18b, bairdNJP19, liPRA20, huXXX20}. 

Despite the great interest and the number of potential applications of the ultrarelativistic particle beam-laser pulse colliding scheme, only recently this kind of experiments have provided, for example, the first evidence of RR signatures in the collision of an ultrarelativistic electron beam with an intense laser pulse~\cite{colePRX18, poderPRX18}. However, these experiments were strongly limited by the uncertainty of the laser pulse and electron beam parameters at collision, which is also determined by shot-to-shot fluctuations. Indeed, these uncertainties did not allow to conclusively discriminate among different radiation emission models~\cite{colePRX18, poderPRX18}. In particular, the stochastic nature of RR, which in the quantum regime is the result of multiple incoherent photon emissions~\cite{dipiazzaPRL10}, still needs to be experimentally demonstrated~\cite{colePRX18, poderPRX18, liPRA18b, liPRA20, huXXX20}. Given that even the modeling of the basic SFQED process of single photon emission by an electron in the presence of arbitrary background fields is still under active theoretical investigation~\cite{dipiazzaPRA18, ildertonPRA19, dipiazzaPRA19, raicherXXX20}, an experimental validation of SFQED models at least in the relatively simple case of a laser pulse becomes imperative.

Here we demonstrate that, in the collision of an ultrarelativistic electron beam with a counterpropagating laser pulse, the stochastic nature of quantum emissions leads to an asymmetry in the electron momentum distribution transverse to electron beam propagation direction. After some centimeters propagation, this induced asymmetry in the momentum distribution is reflected in the electron beam transverse spatial distribution, which can be measured relatively easily by placing a detector plate orthogonal to the beam propagation direction. More importantly, we show that the induced asymmetry is strongly sensitive to the electromagnetic field experienced by the beam particles, therefore providing accurate on-shot information of the electron beam-laser pulse interaction, which is critical for benchmarking theoretical models. Indeed, for example, we show that even 1~$\mu$m electron beam-laser offset has a clear impact on the final transverse electron distribution.

In the following we consider the experimentally relevant regime where $\gamma \gg \xi$ throughout the electron beam-laser pulse interaction. Here $\xi=|e| E_0/m_e \omega c$ is the normalized laser pulse amplitude with $E_0$ being the peak laser field and $\omega$ its central angular frequency, while $\gamma$ is the relativistic factor of the beam electrons. As it will be clear, in contrast to the electron beam energy spread~\cite{neitzPRL13} or transverse electron beam spreading\cite{greenPRL14}, the RR induced transverse asymmetry has clearly measurable effects even when the quality of the initial electron beam energy spectrum is low, as typically occurs in all-optical experiments~\cite{colePRX18, poderPRX18}. In fact, in this case the large initial electron beam energy spread does not significantly change during the interaction, and the transverse spreading is modest when $\gamma$ remains much larger than $\xi$. 

We start considering an electron propagating along the positive $z$ axis and colliding head-on with a plane-wave pulse linearly polarized along the $x$ axis with electric field $E_x(\varphi)=\xi f(\varphi) \sin(\varphi+\delta)$ normalized to $m_e \omega c/|e|$,  
where $f(\varphi)\leq1$ is a smooth temporal envelope, $\varphi$ the pulse phase, $\delta$ a constant phase, and $\int_{-\infty}^{\infty}{E_x(\varphi)\,d\varphi}=0$. Notably, an analytic solution for this problem exists in the classical regime, where RR effects are described by the Landau-Lifshitz equation~\cite{dipiazzaLMP08}. The final electron momentum after the interaction with a quasimonochromatic plane-wave pulse 
is~\cite{tamburiniPRE14}
\begin{equation}
\label{classRR}
\bm{p}_f \approx \frac{\bm{p}_i}{1 + r_{R} \rho_i \int_{-\infty}^{\infty}{E_x^2(\varphi)\,d\varphi}},
\end{equation}
where $\rho_i = \gamma_i (1 + p_{zi}/\gamma_i m_e  c)$ is the initial electron Doppler factor, $r_{R} = 4\pi e^2 / 3 m_e c^2 \lambda \approx 1.18 \times 10^{-8} / \lambda_{\mu\text{m}}$, $\lambda=2\pi c /\omega$ is the pulse wavelength, and where the subscripts $i$ and $f$ denote the initial and final value of the corresponding quantity, respectively. 
Thus, in the classical realm not only the momentum space contracts as predicted for any arbitrary field configuration~\cite{tamburiniNIMA11}, but from \eqref{classRR} in the interaction with a plane-wave pulse the final momentum contraction is nearly isotropic. Consequently, an electron beam with initial cylindrical symmetry around its propagation axis, such as those typically used in all-optical electron beam-laser pulse experiments~\cite{colePRX18, poderPRX18}, preserves its symmetry after interaction. Note that the initial cylindrical symmetry of the electron beam is broken \emph{during} the electron beam-pulse interaction, even with negligible RR effects. This results into radiation being emitted with a characteristic asymmetric pattern in the plane perpendicular to the electron beam propagation direction, which can be used to infer the peak laser pulse intensity~\cite{har-shemeshOP12, blackburnXXX20}.

Since \eqref{classRR} holds in the classical regime where $\chi \ll 1$, the question of quantum effects naturally comes up. Quantum corrections to the classical results arise both because classically no upper bound in the frequency of the emitted radiation exists, which results into an overestimate of energy losses, and because in QED emission is a stochastic process rather than a continuous energy loss. The overestimate of energy losses can be corrected phenomenologically by multiplying the classical radiation reaction force by a weighting function $g(\chi)=I_{\text{Q}}/I_{C}$, where~\cite{kirkPPCF09, baier-book}
\begin{equation}
I_{\text{Q}} = \frac{e^2 m_e^2}{3 \sqrt{3} \pi \hbar^2} \int_0^{\infty}{\frac{u(4u^2+5u+4)}{(1+u)^4} \text{K}_{2/3}\left(\frac{2u}{3\chi}\right)du}
\end{equation}
is the quantum radiation intensity, with $\text{K}_{n}$ being the modified Bessel function of the second kind, and $I_{C}$ = $2 e^2 m_e^2 \chi^2 / 3 \hbar^2$ is the classical radiation intensity. This correction leaves the electron beam-pulse dynamics qualitatively unaffected and only quantitatively corrects the prediction of \eqref{classRR}. In fact, one can show that in this quantum-corrected continuous RR model the final electron momentum is obtained by replacing $E_x^2(\varphi)$ in the integrand at the denominator of \eqref{classRR} with $g[\chi(\varphi)] E_x^2(\varphi)$. On the contrary, stochastic instead of continuous emission of radiation results into qualitatively new features. In particular, it leads to the competition between a phase-space contraction tendency, which dominates in the classical regime, and an opposite tendency to diffusion peculiar to the quantum regime~\cite{neitzPRL13, vranicNJP16}.

\begin{figure}[tb]
\centering
\includegraphics[width=\linewidth]{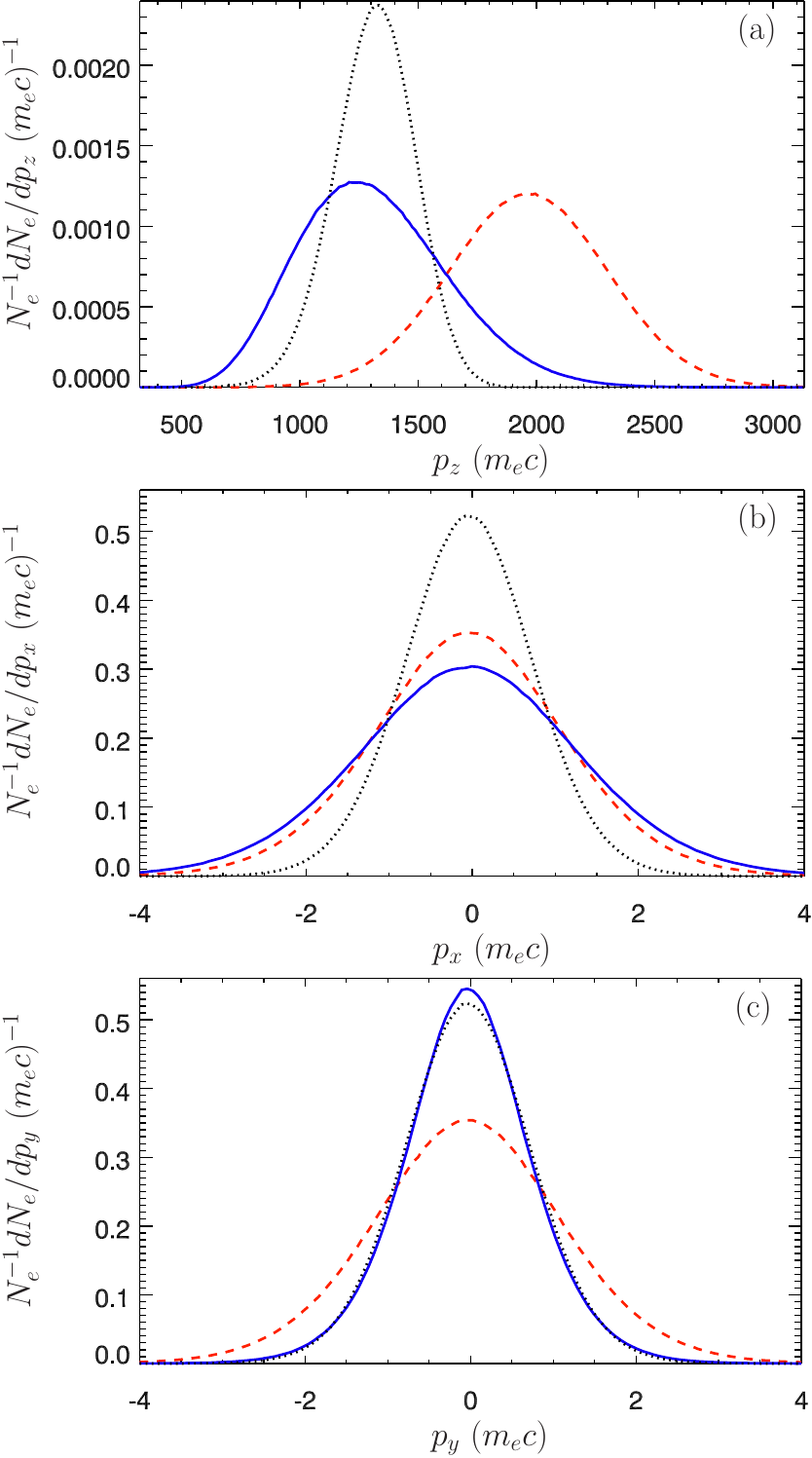} 
\caption{Electron beam momentum distribution. The initial electron beam distribution (red dashed line), the quantum-corrected continuous RR model (black dotted line), the stochastic RR model (solid blue line).  The electron beam propagates along $z$ and collides head-on with a 45~fs FWHM plane-wave pulse with $\xi=15$ linearly polarized along $x$.}
\label{fig:1}
\end{figure}

We now consider the electron dynamics in the case of stochastic photon emissions. We start analyzing the $p_y$ component of the momentum, i.e., the transverse component of the electron momentum perpendicular to the laser pulse polarization axis $x$. Since in the interaction with a plane-wave pulse the Lorentz force along $y$ is zero, $p_y$ changes only because of photon emissions. Thus, by considering the regime $\gamma \gg \xi \gg 1$ and by using the collinear emission approximation, after $N$ photon emissions the electron momentum along $y$ is
\begin{equation}
\label{eq:RRpy}
p_{y,f} = p_{y,i} \displaystyle\prod_{j=1}^{N} \left(1-\frac{\varepsilon_{\gamma,j}}{\varepsilon_{e,j-1}} \right),
\end{equation}
where $\varepsilon_{e,0}$ is the initial electron energy, while $\varepsilon_{e,j}$ and $\varepsilon_{\gamma,j}$ are the electron energy after $j$ emissions and the emitted photon energy at the $j^{\text{th}}$-emission, respectively. From \eqref{eq:RRpy}, $p_{y,f}$ systematically decreases at each emission because $\varepsilon_{\gamma,j} < \varepsilon_{e,j-1}$. This implies a stochastic RR induced momentum decrease similar to the classical and semiclassical RR cases. In practice, $p_{y,f}$ diminishes until approximately $\sim m_e c/\gamma$, which is the lower bound determined by the fact that each photon emission actually occurs within a cone with $\sim1/\gamma$ aperture along the instantaneous electron propagation direction~\cite{baier-book}. A similar analysis can be carried out for the momentum along the plane-wave pulse polarization axis $x$, which gives
\begin{align}
\label{eq:RRpx}
p_{x,f} = & p_{x,i} \displaystyle\prod_{j=1}^{N} \left(1-\frac{\varepsilon_{\gamma,j}}{\varepsilon_{e,j-1}} \right) \nonumber \\
-& m_e c \Bigg[ \displaystyle\sum_{k=1}^{N} \int_{\varphi_{k-1}}^{\varphi_{k}}{d\phi \, E_x(\phi) \displaystyle\prod_{j=k}^{N}
\left(1-\frac{\varepsilon_{\gamma,j}}{\varepsilon_{e,j-1}} \right)} \nonumber \\
+& \int_{\varphi_{N}}^{\infty}{d\phi \, E_x(\phi)} \Bigg],
\end{align}
where $\varphi_0=-\infty$ is the initial electron phase, and $\varphi_k$ is the electron phase at which the $k^{\text{th}}$-photon emission occurs. Both in \eqref{eq:RRpy} and in \eqref{eq:RRpx} the effect of the photon recoil is to reduce the electron momentum at the emission, but for $p_x$ the laser pulse performs a work between successive emission events. Thus, $p_x$ can increase up to $m_e c \xi$, which is possible, for example, for a single photon emission occurring in the central region of the pulse and taking away almost all the electron energy  [see \eqref{eq:RRpx}]. However, for a quasimonochromatic pulse the net transverse momentum gain typically remains modest and much smaller than $m_e c \xi$ (see below).

\begin{figure}[tb]
\centering
\includegraphics[width=\linewidth]{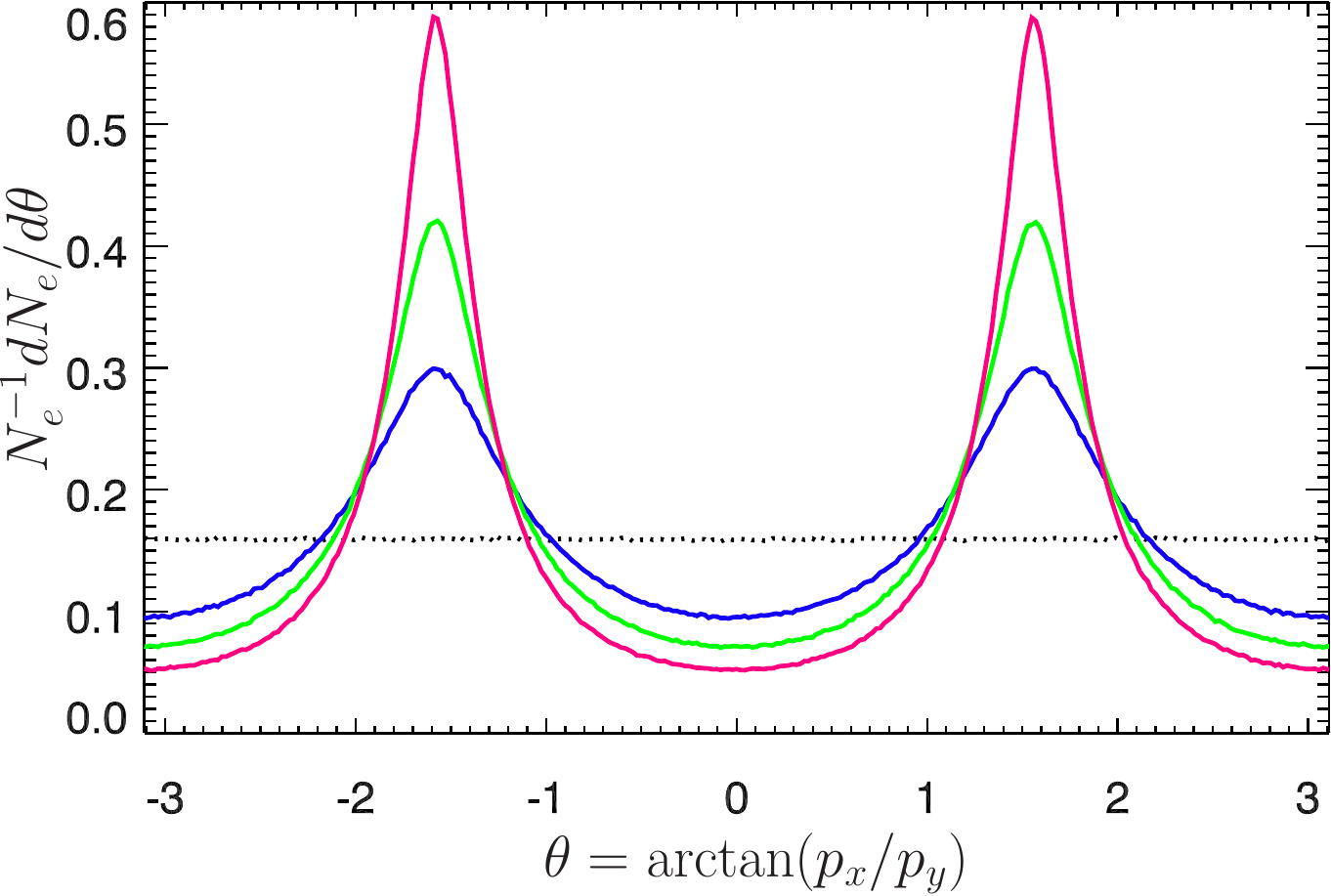} 
\caption{Electron beam azimuthal angle $\theta$ distribution. The initial (black dotted line) and the final distribution with stochastic RR in the head-on collision with a 45~fs FWHM plane-wave pulse linearly polarized along $x$ with $\xi=15$ (blue line), $\xi=18$ (green line), and $\xi=21$ (magenta line).}
\label{fig:2}
\end{figure}
\begin{figure}[h]
\centering
\includegraphics[width=\linewidth]{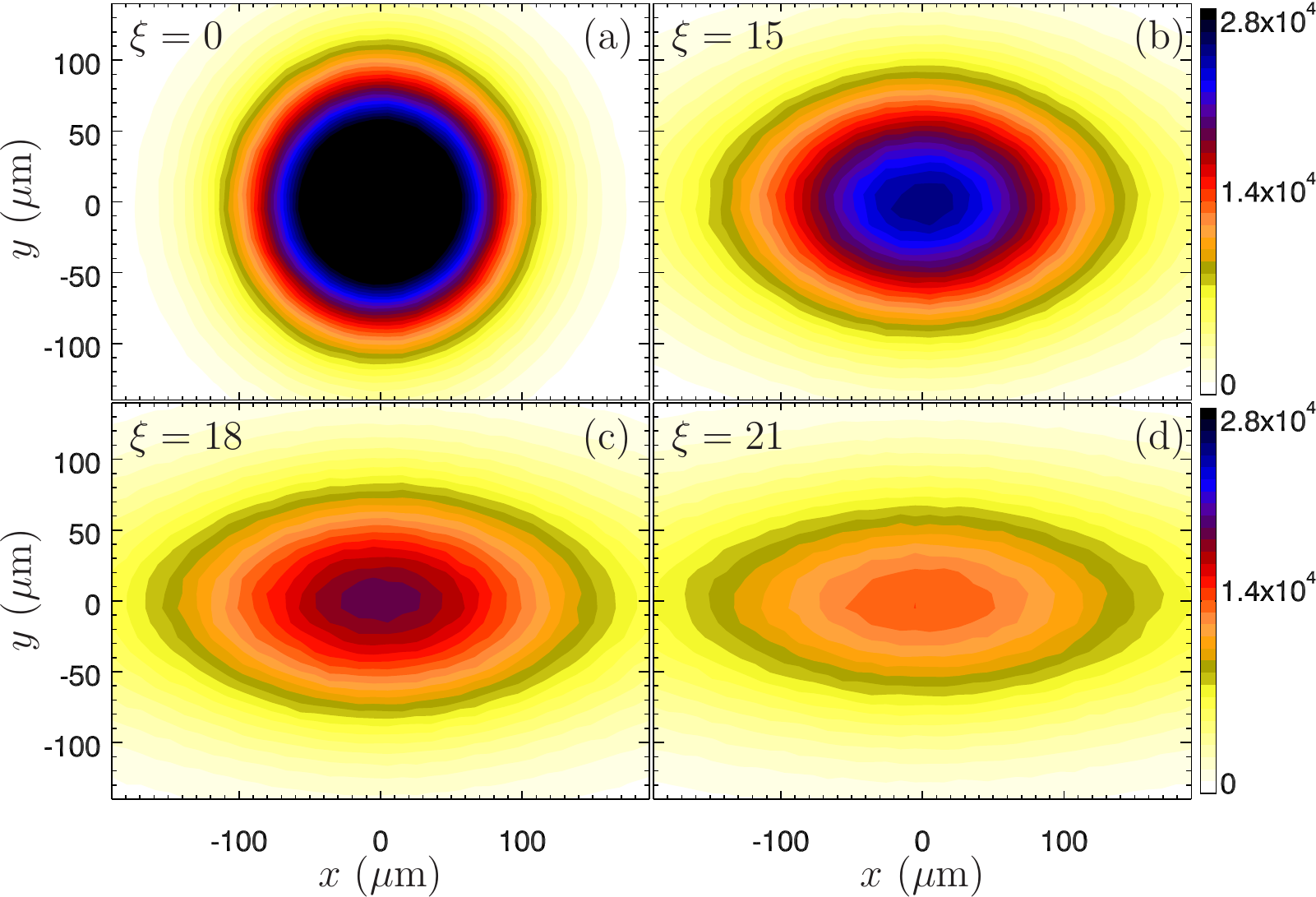} 
\caption{Electron beam transverse spatial distribution after the collision with a 45~fs plane-wave pulse linearly polarized along the $x$ axis followed by 10~cm ballistic propagation. (a) no interaction, (b) $\xi=15$, (c) $\xi=18$, (d) $\xi=21$.}
\label{fig:3}
\end{figure}
\begin{figure*}[tb]
\centering
\includegraphics[width=\linewidth]{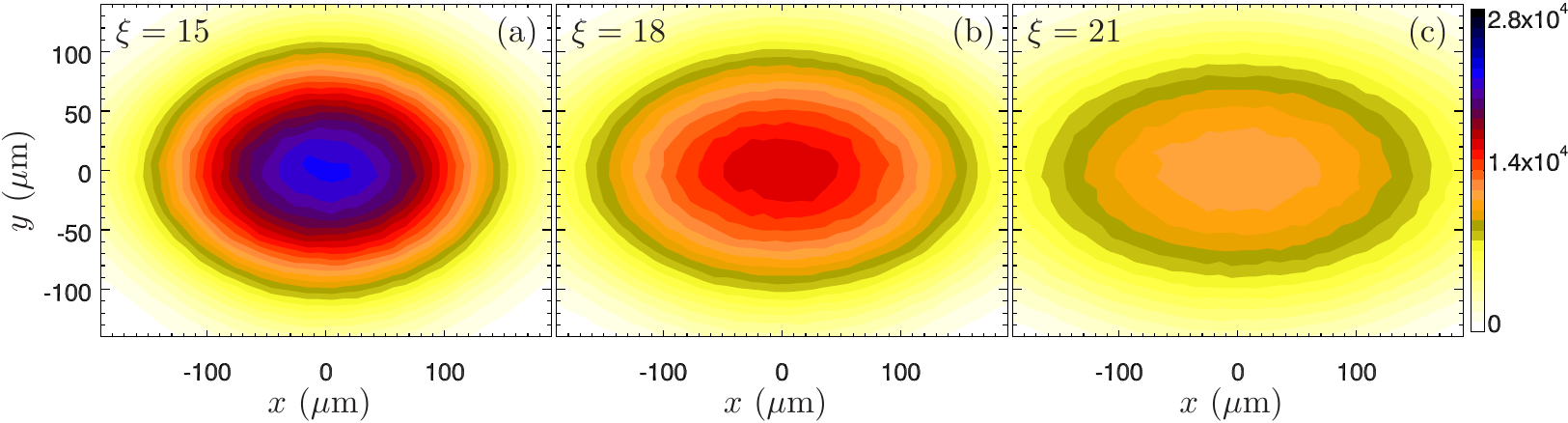} 
\caption{Electron beam transverse spatial distribution after the collision with a 45~fs focused Gaussian laser pulse with $w_0 = 4$~$\mu$m waist radius linearly polarized along $x$ followed by 10~cm ballistic propagation. (a) $\xi=15$, (b) $\xi=18$, (c) $\xi=21$.}
\label{fig:4}
\end{figure*}
\begin{figure}[h]
\centering
\includegraphics[width=\linewidth]{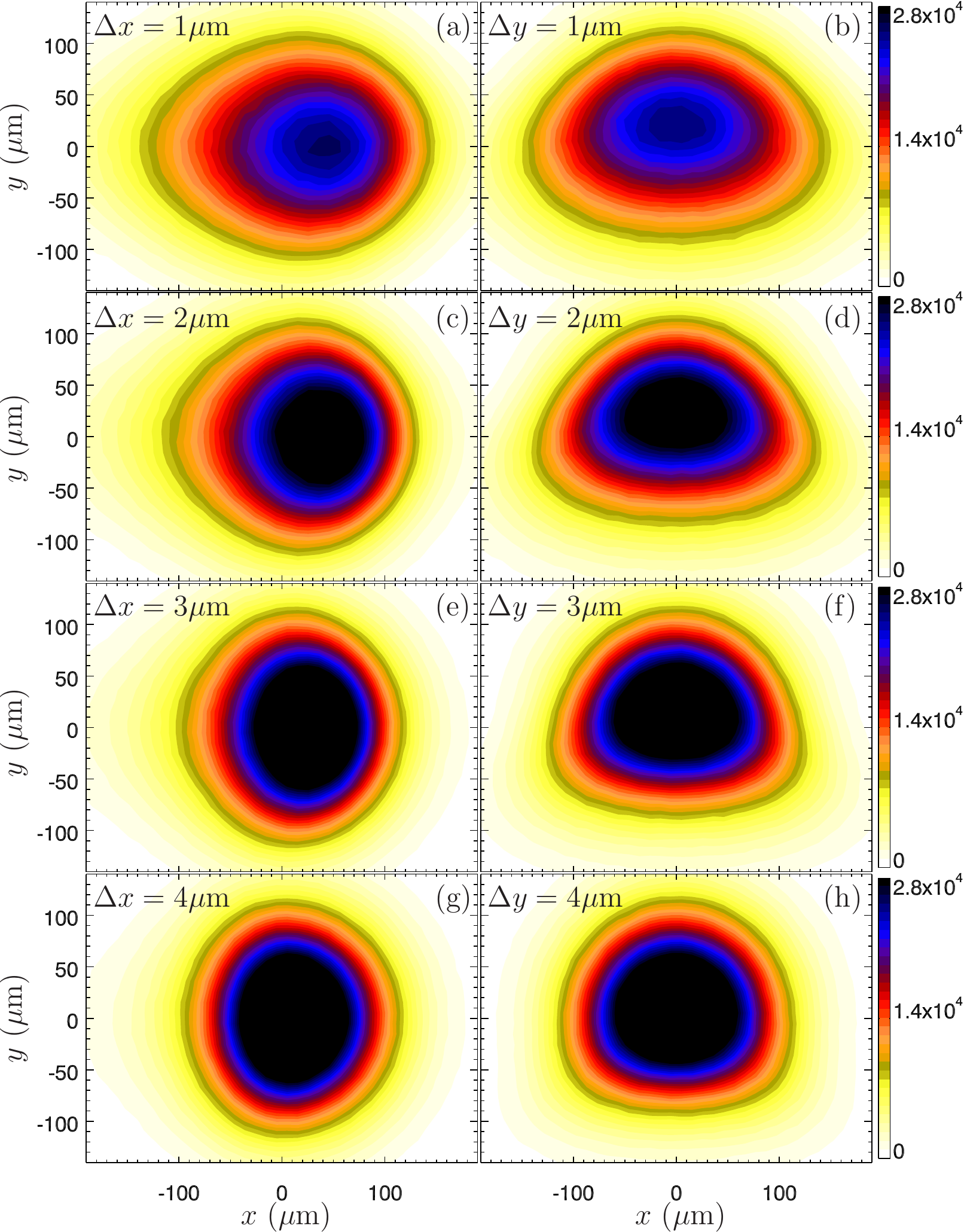} 
\caption{Electron beam transverse spatial distribution after the collision with a 45~fs focused laser pulse with $w_0 = 4$~$\mu$m and $\xi=15$ linearly polarized along $x$ followed by 10~cm ballistic propagation. Left column, 1~$\mu$m (a), 2~$\mu$m (c), 3~$\mu$m (e), 4~$\mu$m (g) misalignment along $x$. Right column, 1~$\mu$m (b), 2~$\mu$m (d), 3~$\mu$m (f), 4~$\mu$m (h) misalignment along $y$.}
\label{fig:5}
\end{figure}

Figure~\ref{fig:1} shows the initial (red dashed line) and final electron beam momentum distribution as predicted by the quantum-corrected continuous RR model (black dotted line) and by the stochastic quantum RR model (solid blue line) in the head-on collision with a 45~fs duration full width at half maximum (FWHM) of the intensity Gaussian plane-wave pulse propagating along $z$ with $\xi=15$ amplitude, $\lambda=0.8$~$\mu$m wavelength, and linearly polarized along $x$. Initially, the electron beam has cylindrical symmetry around its propagation axis, and its energy distribution is Gaussian with 1~GeV mean energy and 400~MeV energy spread FWHM. The initial beam spatial distribution is Gaussian with 12~$\mu$m length FWHM, 2.8~$\mu$m transverse size FWHM, and 1.4~mrad FWHM angular aperture, while the beam charge is 1.6~pC. Figure~\ref{fig:1}(a) shows that the quantum-corrected continuous RR model and the stochastic RR model predict nearly the same final mean longitudinal momentum. However, the stochastic RR model predicts a longitudinal momentum spread comparable to that of the initial beam, whereas the momentum spread is significantly reduced in the continuous RR model [see Fig.~\ref{fig:1}(a)]. The same conclusion applies to the electron momentum distribution along the pulse polarization axis $x$, which presents a modest width increase in the stochastic RR model and a significant width reduction in the continuous RR model [see Fig.~\ref{fig:1}(b)]. In sharp contrast, almost the same final distribution is obtained in the quantum-corrected and continuous RR models for the transverse momentum orthogonal to the pulse polarization axis [see Fig.~\ref{fig:1}(c)]. Thus, a peculiar manifestation of stochastic RR is the induced electron beam asymmetry in the plane perpendicular to the pulse propagation direction, which is absent in continuous emission models.

Figure~\ref{fig:2} reports the final electron beam azimuthal angle $\theta = \arctan(p_x/p_y)$ distribution for the electron beam collision with a plane wave pulse with the same parameters as reported above but for field amplitude $\xi=15$ (blue line), $\xi=18$ (green line), and $\xi=21$ (magenta line). The strong dependence of the azimuthal angle distribution on the normalized field amplitude $\xi$ is evident in Fig.~\ref{fig:2}. Moreover, this induced transverse angular asymmetry is later reflected in the spatial electron beam distribution. In fact, for sufficiently long propagation distance, the transverse size of the electron beam is linearly proportional to the transverse momentum as the initial size of the beam becomes negligible. Thus, one can introduce an asymmetry parameter $\mathcal{A}=(\sigma_x - \sigma_y)/(\sigma_x + \sigma_y)$ directly proportional to the momentum asymmetry, where $\sigma_x \approx p_{x,f} d_z/\gamma_f m_e c$ and $\sigma_y \approx p_{y,f} d_z/\gamma_f m_e c$ are the beam width along $x$ and $y$, respectively, with $d_z$ being the ballistic propagation distance.

Figure~\ref{fig:3} displays the transverse spatial electron beam distribution for the case where no colliding pulse is present [Fig.~\ref{fig:3}(a)] and for the three cases of Figure~\ref{fig:2} after 10~cm ballistic propagation. By increasing $\xi$, the electron beam width increases along the polarization axis $x$ and simultaneously reduces along $y$ (see Fig.~\ref{fig:3}). Note that $\sigma_y$ significantly decreases with increasing $\xi$ due to the cumulative effect of photon emissions on $p_y$, whereas $\sigma_x$ is weakly affected. Similar results are obtained in the electron beam collision with a focused laser pulse. Figure~\ref{fig:4} reports the simulation results obtained with the same parameters as in Fig.~\ref{fig:3} but for a focused Gaussian laser pulse with $w_0=4$~$\mu$m waist radius. The reduction of the transverse compression along $y$ observed in Fig.~\ref{fig:4} as compared to the corresponding plane-wave pulse case reported in Fig.~\ref{fig:3} originates from the strong sensitivity of the stochastic RR induced transverse asymmetry effect on the field strength experienced by beam electrons. In fact, when the electron beam transverse size at collision is comparable or larger than the laser pulse waist radius, electrons in the periphery experience smaller fields than in the central region.

Finally, we consider the possible presence of a small misalignment between the electron beam and the focused laser pulse propagation direction. Figure~\ref{fig:5} shows the results obtained with the same parameters as in Fig.~\ref{fig:4}(a) but with a transverse misalignment ranging from 1 to 4 $\mu$m either along the laser pulse polarization axis $x$ (left column of Fig.~\ref{fig:5}) or along its orthogonal direction $y$  (right column of Fig.~\ref{fig:5}). The presence of even a small misalignment determines a pronounced left-right asymmetry if the misalignment is along $x$ (see the left column of Fig.~\ref{fig:5}) and a characteristic up-down asymmetry if the misalignment is along $y$ (see the right column of Fig.~\ref{fig:5}). Thus, the transverse electron beam spatial distribution after collision provides information both on the strength of the laser pulse field experienced by beam electrons and on the laser pulse-electron beam overlap at interaction. We stress that, for the considered case of ultrarelativistic electron beam collision with $\gamma \gg \xi$ throughout the interaction, the transverse electron beam spatial deformation both in the collinear case (see Fig.~\ref{fig:4}) and in the presence of a misalignment (see Fig.~\ref{fig:5}) is determined by stochastic RR effects. Indeed, no significant transverse deformation is observed in simulations with the same parameters but with continuous RR models. 

\begin{acknowledgments}
We thank Antonino Di Piazza, Karen Z. Hatsagortsyan and Christoph H. Keitel for useful discussions. As this manuscript was being prepared, Ref.~\cite{huXXX20} appeared on the arXiv, with some overlap in noticing the transverse angular asymmetry induced by stochastic RR. No overlap exists in the electron beam-laser pulse diagnostic, which is the central point of the present work.
\end{acknowledgments}

\bibliography{My_Bibliography}

\end{document}